\documentclass[twocolumn,aps,prl,superscriptaddress]{revtex4-2}

\usepackage{amsmath}
\usepackage{graphicx}
\usepackage{MnSymbol}
\usepackage{parskip}
\usepackage{xcolor}
\usepackage{makecell}
\usepackage{booktabs}
\usepackage{multirow}

\usepackage{comment}

\begin{document}

\title{Surface separation in elastoplastic contacts}

\author{A. Almqvist}
\affiliation{Division of Machine Elements, Lule{\aa} University of Technology, 97187, Lule{\aa}, Sweden}

\author{B.N.J. Persson}
\affiliation{Peter Gr\"unberg Institute (PGI-1), Forschungszentrum J\"ulich, 52425, J\"ulich, Germany}
\affiliation{State Key Laboratory of Solid Lubrication, Lanzhou Institute of Chemical Physics, Chinese Academy of Sciences, 730000 Lanzhou, China}
\affiliation{MultiscaleConsulting, Wolfshovener str. 2, 52428 J\"ulich, Germany}

\begin{abstract}
Understanding the contact between rough surfaces undergoing plastic deformation is crucial in many applications. 
We study the effect of plastic deformation on the surface separation between two solids with random roughness. 
Assuming a constant penetration hardness, we propose an iterative smoothing procedure (akin to elastoplastic ``shakedown'') within Persson's multiscale contact mechanics theory to obtain the average surface 
separation by applying the elastic formulation to an effective power spectrum that accounts for plastic smoothing.
Deterministic numerical simulations based on the boundary element method are used to validate the procedure 
and show good agreement with the theoretical predictions. 
The treatment also provides a route to incorporate plastic stiffening of the roughness as the stress state becomes increasingly hydrostatic at large plastic deformation.
\end{abstract}

\maketitle

\setcounter{page}{1}
\pagenumbering{arabic}




{\bf 1 Introduction} 

All solid bodies have rough surfaces. When two solids are pressed together, only the highest asperities carry the load, and even under high loading conditions, there will be contact only at a small fraction of the nominal area. In many cases, the local stresses in these asperity contacts are sufficiently high that the solids yield plastically, at least in part of the contact region. Predicting the real contact area, the distribution of contact stresses, and the interfacial separation in the non-contacting regions is therefore a central problem in elastoplastic contact mechanics with relevance to a wide range of practical applications.

One analytical approach to the elastoplastic contact between solids is the Persson contact mechanics theory. This theory, in its simplest form, assumes that the solids yield plastically when the local stress reaches the penetration hardness $\sigma_{\rm P}$. The penetration hardness is determined in indentation experiments, where a very hard object (often diamond) of well-defined shape (usually a ball or pyramid) is pressed into contact with the solid. 
%
The penetration hardness, $\sigma_{\rm P}$, is then obtained as the ratio between the applied load and the projected area of the plastically deformed indentation.
The elastoplastic contact mechanics theory has been applied to the leakage of metallic seals \cite{Metal} and to the leakage of syringes with Teflon laminated rubber stoppers \cite{Nest}. In a recent study, Lambert and Brodsky~\cite{Brodsky} have applied the theory (with a length-scale dependent penetration hardness) to the contact between the surfaces in earthquake faults.

In \cite{PRE}, we demonstrated that the predictions of the Persson contact mechanics theory, for the contact area and the contact stress distribution for elastoplastic contacts, are in good agreement with deterministic numerical simulations based on the boundary element method. Here, we present some additional results for the pressure dependency of the elastic and plastic contact area. The main aim of this study is, however, to show that the Persson theory, with an effective surface roughness power spectrum for the plastically deformed surface, can also be used to calculate the interfacial separation between surfaces in elastoplastic contact. We propose that this treatment may allow for inclusion of the plastic stiffening of the roughness as the stress state becomes hydrostatic at large plastic deformation, see Refs.~\cite{A23, A24, A25, A28, A29, A30}. We note that numerical results for the average surface separation during loading and unloading have recently been presented by R\`afols et al.~\cite{perezrafols2025general}, but no comparison to theoretical results was given.

\begin{figure}
	\includegraphics[width=0.35\textwidth,angle=0.0]{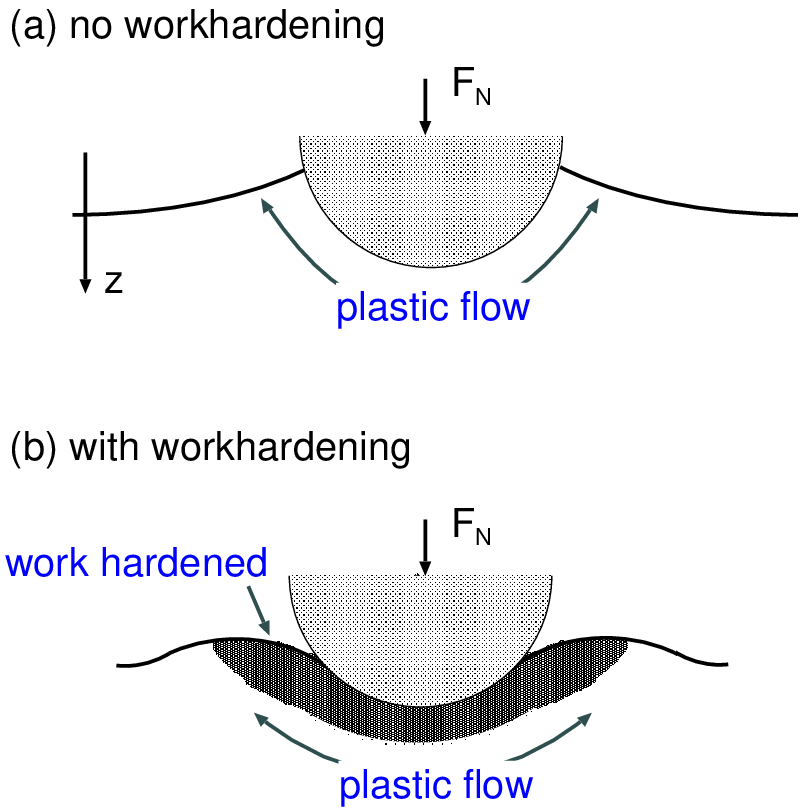}
	\caption{\label{WithoutAndWithWorkhardening.eps}
(a) The deformation and flow of metal around an indenter for a work-hardened specimen. For a rigid indenter there is a raised ridge or pile-up of material near the edge of the indentation. (b) The deformation and flow of metal around an indenter for an annealed specimen. In this case the indenter is sinking-in near the edge of the indentation. The penetration hardness $\sigma_{\rm P} = F_{\rm N}/A$, where $F_{\rm N}$ is the applied normal force and $A$, is the projection of the indentation area on the horizontal $xy$-plane.
}
\end{figure}

\vspace{0.3cm}
{\bf 2 Plastic flow in contact mechanics}

Plastic flow at interfaces between solids is a complex topic, but much information has been gained from hardness measurements. Here we briefly review some of the most important results. For a detailed study see Tabor~\cite{Tabor}.



Consider a rigid asperity indenting a metal that yields plastically. It is well known that hydrostatic pressure does not produce plastic flow, whereas plastic flow is induced by the shear stress. For crystalline materials this implies that plastic deformation is always accompanied by slip of atomic planes over one another. A full analysis of the indentation process shows that nearly two-thirds of the mean contact pressure is hydrostatic and therefore plays no part in producing plastic flow, while only the remaining one-third is effective in driving plastic deformation. This gives the classical Tabor relation $$\sigma_{\rm P} \approx 3 \sigma_{\rm Y},$$ between the penetration hardness, $\sigma_{\rm P}$, and the yield stress in elongation, $\sigma_{\rm Y}$, with the exact prefactor depending somewhat on work hardening, indenter geometry, and friction.


For a rigid–perfectly plastic, incompressible solid with a constant yield (flow) stress, material displaced by an indenter flows laterally and upward along the indenter faces, producing a pile-up ridge around the contact perimeter. The same qualitative behavior occurs in metals that are already heavily work-hardened and exhibit negligible additional work-hardening over the indentation-strain range. When interfacial friction is non-negligible, this pile-up is most pronounced for weak or vanishing strain hardening. Figure~\ref{WithoutAndWithWorkhardening.eps}(a) shows the characteristic pile-up at the contact perimeter.

With highly annealed metals (i.e., those that exhibit pronounced strain hardening), the behavior is different. When the indenter first begins to sink into the metal, the material adjacent to the indenter undergoes large plastic strains and work-hardens relative to the undeformed metal farther away. Hence, when the indenter sinks in farther, it carries the surrounding metal with it, thereby acting as an enlarged indenter deforming the metal adjacent to it. The result is that the displaced metal always appears to be moving out farther and farther away from the indenter itself. This leads to a depression of the metal immediately adjacent to the indenter and a slight piling-up some distance away, as illustrated in Fig.~\ref{WithoutAndWithWorkhardening.eps}(b).

This discussion demonstrates that plastic flow in the contact between solids depends on both material properties and surface preparation.
In what follows, we will not include these complexities but consider more idealized plastic flow laws.

\begin{figure}
\includegraphics[width=0.25\textwidth,angle=0.0]{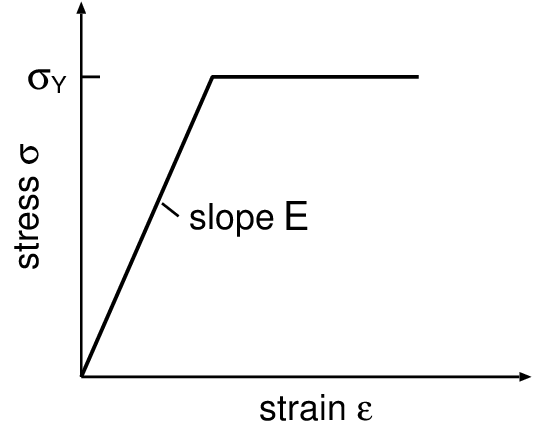}
\caption{\label{ElastoPlasticStressStrain.eps}
The relation between stress and strain in uniaxial tension for the simplest elastoplastic material model. The material deforms elastically with slope $E$ up to the yield stress $\sigma_{\rm Y}$, beyond which it flows plastically without strain hardening.
}
\end{figure}

\vspace{0.3cm}
{\bf 3 Plastic deformation: Contact area} 

When the stress in the asperity contact region becomes high enough, plastic flow occurs. 
In the simplest model, it is assumed that in elongation a material deforms as a linear elastic solid until the stress reaches a critical level, the so-called plastic yield stress, $\sigma_{\rm Y}$, where it flows without strain hardening, see Fig.~\ref{ElastoPlasticStressStrain.eps}. 
We note that $\sigma_{\rm Y}$ and $\sigma_{\rm P} \approx 3 \sigma_{\rm Y}$ of materials often depend on the length scale (or magnification), which in principle can be included in the formalism we use, see Refs.~\cite{Preview,Brodsky}.

All solid surfaces exhibit roughness over many length scales. When a surface is observed at magnification $\zeta$, only the roughness components with wavelengths $\lambda>L/\zeta$, where $L$ is a reference length (such as the linear size of the surface), can be observed. In the following, we will use the wavenumber $q=2\pi/\lambda$, instead of the wavelength, and we define $q_0=2\pi /L$ so that $q=\zeta q_0$.

The Persson contact mechanics theory is based on studying the probability distribution $P(\sigma,\zeta)$ of contact stress $\sigma$ as the magnification $\zeta$ increases. 
When calculating $P(\sigma,\zeta)$, only the roughness components with wavenumber $q< q_0 \zeta$ are included.
This function obeys a diffusion-like equation where time is replaced by magnification $\zeta$ and the spatial coordinate by the stress $\sigma$, and where the diffusivity depends on the surface roughness power spectrum and on the elastic properties of the solids. 
To solve the equation, suitable boundary conditions must be imposed. The influence of plastic flow is incorporated by requiring $P(0,\zeta) = 0$ with $P(\sigma_{\rm P},\zeta) = 0$. In \cite{PRE}, we showed that these boundary conditions naturally emerge from the numerical model described in Sec.~7.

Assume that an elastoplastic block with surface roughness is squeezed against a flat rigid surface with the nominal contact pressure $\sigma_0 = F_{\rm N}/A_0$, where $F_{\rm N}$ is the applied normal force and $A_0$ is the nominal contact area.
The fractions of the nominal contact area occupied by elastic and plastic contacts, denoted by $A_{\rm el} (\zeta) /A_0$ and $A_{\rm pl} (\zeta) /A_0$, respectively, when the interface is observed at magnification $\zeta$, are given by 
$$
\frac{A_{\rm el} (\zeta )}{A_0}  = \frac{2}{\pi} \sum_{n=1}^\infty \, \frac{1}{n} \left[ 1-(-1)^n\right ] {\rm sin} (s_n\sigma_0) \, e^{-s_n^2 G(\zeta)}, \eqno(1)
$$
and
$$
\frac{A_{\rm pl} (\zeta )}{A_0} = \frac{\sigma_0}{\sigma_{\rm P}}+\frac{2}{\pi} \sum_{n=1}^\infty \, \frac{(-1)^n}{n} {\rm sin} (s_n\sigma_0) \, e^{-s_n^2 G(\zeta)}, \eqno(2)
$$
respectively, where $s_n = n \pi/\sigma_{\rm P}$, and where $\sigma_0$ is the applied squeezing pressure. The function $G(\zeta)$, appearing in these expressions, is defined as 
$$ 
G (\zeta) =\frac{\pi}{4} (E^*)^2 \int_{q_0}^{\zeta q_0} dq \ q^3 C(q)S(q) \eqno(3)
$$
where $E^* = E/(1-\nu^2)$ is an effective elastic modulus, $C(q)$ is the surface roughness power spectrum, and $S(q)$ is the ``correction factor'':
$$ 
S(q)\approx \gamma +(1-\gamma) P(\zeta )
$$ 
where $P(\zeta) = A(\zeta)/A_0$ with $A(\zeta) =A_{\rm el}(\zeta)+A_{\rm pl} (\zeta )$ (see Ref.~\cite{ep11}).
Physically, $G(\zeta)$ can be viewed as a measure of how the surface roughness at different length scales amplifies local stress variations -- the finer the asperities included at higher magnification, the larger the fluctuations in contact stresses become. 

In (3) $S(q)$ is an elastic-energy reduction factor that accounts for the partial relaxation of interfacial stresses due to incomplete contact. 
The factor $\gamma$ was originally determined by comparing the average surface separation, $\bar{u}(p)$, predicted by the theory with numerical simulation results, and has been found to be (see Ref.~\cite{ep11}) $\gamma\approx 0.5$. Here we will use the same value as used in \cite{PRE}, namely $\gamma=0.536$. 
For $A/A_0 << 1$ we get $S\approx \gamma$ and this limit is equivalent to setting $S=1$ and replacing the effective elastic modulus $E^{*}$ by $\approx E^{*} \surd \gamma \approx 0.73 E^{*}$. 


In what follows, we consider the average surface separation $\bar u$ as a function of the instantaneous squeezing pressure $p$. First, we increase $p$ from $0$ to a prescribed maximum, $\sigma_0$, sufficient to induce plastic yielding and permanently deform the rough surface; then we decrease $p$ from $\sigma_0$ back to $0$, and record $\bar u(p)$ during unloading. At this stage (and any subsequent reloading within $0 \le p \le \sigma_0$), we assume a purely elastic response of the plastically deformed surface profile.

\begin{figure}
\includegraphics[width=0.47\textwidth,angle=0.0]{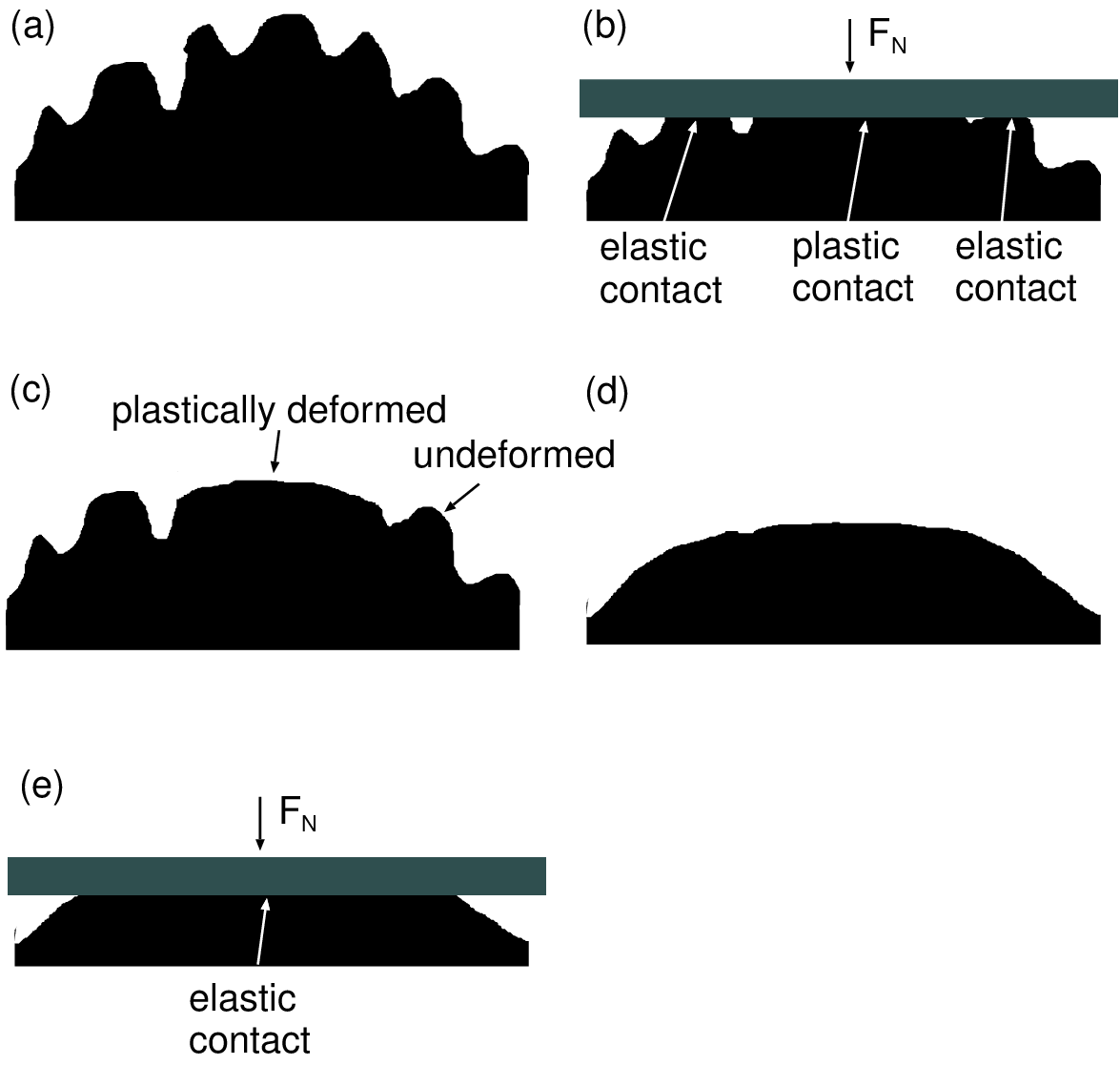}
\caption{\label{plast1.eps}
(a) An asperity on an elastoplastic solid with random roughness. 
(b) A rigid flat plate pressed against the solid. Some of the asperities are elastically deformed, while others yield plastically. 
(c) The solid after removal of the rigid plate. Asperities that deformed elastically have fully recovered their original shape, while those that yielded plastically remain permanently flattened, with partial recovery due to elastic spring-back.
(d) In the theoretical approach where the surface roughness power spectrum is modified, all asperities are smoothed, i.e., also those which were not in contact with the plate. 
(e) When the modified power spectrum is used, the area of real contact will be larger than for the true situation in (b).
}
\end{figure}

\vspace{0.3cm}
{\bf 4 Interfacial separation: qualitative discussion} 

We study the interface between an elastoplastic solid with a randomly rough surface, and a rigid body with a perfectly flat surface at a magnification $\zeta$. At this magnification, we only observe the roughness with wavenumber components below $q=\zeta q_0$, where $q_0$ is the wavenumber of the smallest roughness component. We assume that the solids have nominally flat surfaces and are pressed together with a nominally uniform stress $\sigma_0$. The nominal contact area is denoted by $A_0$. At the magnification $\zeta$, we observe that the fraction $A_{\rm el}(\zeta)/A_0$ of the surface area is in elastic contact, while the fraction $A_{\rm pl}(\zeta)/A_0$ is in plastic contact. For randomly rough surfaces, these quantities can be calculated using (1) and (2).

Here, we will describe how one can estimate how plastic deformation changes the average interfacial separation between the contacting surfaces.  Figure~\ref{plast1.eps}(a) schematically shows a (relatively) large asperity with smaller asperities on an elastoplastic solid, with nominally flat surface. 
If the solid is pressed against a rigid flat surface, some of the asperities deform elastically while others yield plastically. Figure~\ref{plast1.eps}(c)  shows the elastoplastic solid after removing the rigid plate. Some asperities have returned to their original shape, while other asperities are plastically deformed with partial recovery due to elastic spring-back. The net effect is a scale-dependent flattening of the surface roughness profile, which reduces the average surface separation at a given unloading pressure compared with the original surface.

The Persson contact mechanics theory assumes randomly rough surfaces. Even if a surface is randomly rough before plastic deformation, this cannot be the case for the plastically deformed surface. However, one can estimate the average interfacial separation using elastic contact mechanics if the plastically deformed surface is replaced by another (randomly rough) surface where the short wavelength roughness components, which deform plastically, are removed (or reduced) {\it everywhere} as in Fig.~\ref{plast1.eps}(d), i.e., not just close to the top of the big asperities. 

In the theoretical approach that we use to estimate the change in the surface separation, the surface roughness power spectrum is modified by reducing the amplitude of the surface roughness components, which were smoothed plastically during contact with the rigid flat plate in Fig.~\ref{plast1.eps}(b). This corresponds to smoothing the surface everywhere, including the region that was not in contact with the rigid plate. As illustrated in Fig.~\ref{plast1.eps}(e), this will result in a larger elastic contact area than the one shown in Fig.~\ref{plast1.eps}(b). Thus, we use a modified power spectrum to estimate the change in surface separation, while the contact area can be calculated using Persson's contact mechanics theory, which utilizes the power spectrum of the original (undeformed) surface. 

\begin{figure}
\includegraphics[width=0.47\textwidth,angle=0.0]{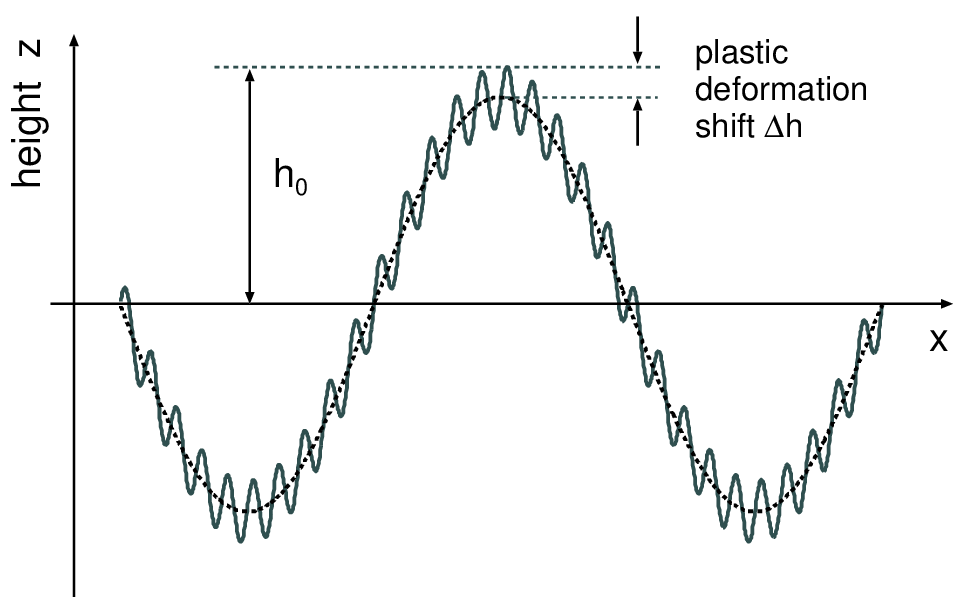}
\caption{\label{1x.2height.sinus.eps}
A surface of an elastoplastic solid with roughness on two length scales. A flat rigid surface will make first contact with the roughness profile at the top of the smaller asperity on top of the larger asperity for $z=h_0$. If the rigid flat is pressed against the elastoplastic solid so that the small asperities (short wavelength roughness) are smoothed close to the top of the larger asperity, then on a second approach to the plastically deformed surface the two solids will make contact at $z=h_0-\Delta h$ where $\Delta h$ is the amplitude of the short wavelength roughness. Thus, the average surface separation at the first contact during the second approach will be $\Delta h$ smaller than for the undeformed surface. The Persson contact mechanics theory assumes randomly rough surfaces, implying that short-wavelength roughness must be smoothed everywhere (dashed line) when used in the theory calculations to estimate the influence of plastic flow on average surface separation.
}
\end{figure}

Figure~\ref{1x.2height.sinus.eps} illustrates the discussion above for an elastoplastic solid with surface roughness on two characteristic length scales. 
A flat rigid surface will first come into contact ($F_{\rm N} = 0^+$) with the roughness profile at the top of the small asperity located on the top of the larger asperity, i.e.~at $z = h_0$. 
If the rigid flat is pressed against the elastoplastic solid so that the small asperities (short-wavelength roughness) undergo plastic flattening near the top of the large asperity, a subsequent approach to the plastically deformed surface will result in first contact ($F_{\rm N} = 0^+$) at a reduced height, $z = h_0 - \Delta h$, where $\Delta h$ denotes the amplitude of the short-wavelength roughness. 
Consequently, the average surface separation at the onset of contact during the second approach will be $\Delta h$ smaller compared with the undeformed surface.

In the Persson contact mechanics theory, the surfaces are assumed to be randomly rough. This assumption is valid only when the short-wavelength roughness has been smoothed over the entire interface. Hence, in the theoretical treatment of plastic flow, this situation corresponds to using the dashed profile in Fig.~\ref{1x.2height.sinus.eps} to represent the modified (smoothed) power spectrum.

\begin{figure}
\includegraphics[width=0.3\textwidth,angle=0.0]{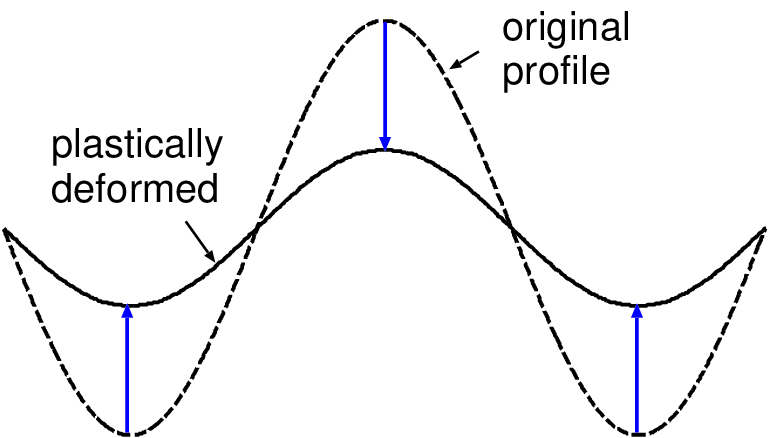}
\caption{\label{1x.2height.sinusVOLUME.eps}
Example of a volume-conserving plastic deformation: When the top of the roughness profile moves downwards, the valley moves upwards.}
\end{figure}

The approach used above to include plastic flow conserved the volume of the elastoplastic solid. Figure~\ref{1x.2height.sinusVOLUME.eps} illustrates the type of volume-conserving plastic deformations implicit in the present theory approach: when the top of the roughness profile moves downwards, the valley moves upwards. In the boundary element simulations presented below, surface elements (represented by grid points) in the contact regions move downwards so that the maximum stress equals the penetration hardness. This numerical approach does not conserve the volume of the elastoplastic solid.

We note that, in reality, it may be nearly impossible to completely flatten a periodic roughness component, see Refs.~\cite{A23, A24, A28, A29, A30}. Thus, Manners~\cite{Manners} used slip-line theory to show that as complete contact was approached, the pressure could grow beyond any bounds. This is due to the stress state becoming hydrostatic and therefore not allowing plastic deformation. In actual contacts, the surface pressure will remain bounded because elastic deformation allows for the surfaces to be pressed together. The average pressure does, however, increase relative to the yield strength and appears to have no upper limit, and also increases with the amplitude of the sinusoidal surface being flattened \cite{Jackson}. 
This effect of plastic strengthening with increasing deformations could, in principle, be accounted for in an approximate way in the theory presented below, by the way we reduce the amplitude of the roughness wavelength components in the smoothing process.



\vspace{0.3cm}
{\bf 5 Influence of plastic flow on the powerspectrum} 
There are several ways to account for the plastic smoothing of a rough surface. 
Let us consider two solids that are pressed into contact until plastic flow occurs and are then separated. 
Experiments indicate that, when such surfaces are repeatedly reloaded at the same relative position and under the same normal force, the first few cycles may involve additional plastic deformation. 
After a number of cycles, however, the plastically modified surface adapts, and the subsequent response becomes purely elastic. This behavior reflects a shakedown process, as described in \cite{frerot_pastewka2024shakedown}, where the contact gradually evolves toward a state in which further loading produces only elastic deformation---albeit of a surface whose effective properties have been altered by prior plasticity. 
The iterative modification of the surface-roughness power spectrum introduced below is intended to mimic this gradual progression toward shakedown.

We generate the power spectrum of the smoothed surface in an iterative manner. At a given magnification $\zeta$, the probability that a point in the contact area belongs to the elastically deformed region is defined as
$$
p_1(q) = \frac{A_{\rm el}(\zeta)}{A_{\rm pl}(\zeta) + A_{\rm el}(\zeta)}, \eqno(4)
$$
where $q=q_0 \zeta$. We now replace the original power spectrum with
$$
C_1 (q) = C(q) p_1(q).\eqno(5)
$$
Using the modified power spectrum, $C_1(q)$, we compute updated values of $A_{\rm el}(\zeta)$ and $A_{\rm pl}(\zeta)$, from which a new probability $p_2(q)$ is obtained using (4). Next, we form the modified power spectrum $C_2(q) = C_1(q) p_2(q)$. This process is repeated $N$ times to form successive modified spectra,
\[
C_2(q) = C_1(q) \, p_2(q), \quad \ldots \quad C_N(q) = C_{N-1}(q) \, p_N(q),
\]
until the plastic contact fraction becomes negligible, i.e.~$A_{\rm pl}(\zeta) \ll A_{\rm el}(\zeta)$. At this stage, we define the plastically modified power spectrum as $C_{\rm pl}(q) = C_N(q)$, which is subsequently used in the Persson elastic contact mechanics theory to calculate the interfacial separation.
 
If desired, the iteration may be truncated at an earlier $N$, where the plastic contact area remains non-negligible compared to $A_{\rm el}(\zeta)$. In this case, one obtains a surface whose roughness components are not fully flattened, representing a state of incomplete shakedown or progressive plastic strengthening of the roughness, as discussed in Sec.~4.

As a limiting case, if $p_1(q)=1$ for $q<q^*$ and $p_1(q)=0$ for $q>q^*$, the iterative procedure above reduces to a sharp cut-off of the power spectrum at $q=q^*$, such that $C_{\rm pl}(q) = 0$ for $q>q^*$. This limiting behaviour is physically consistent, as it corresponds to complete plastic smoothing of all roughness components with wavelengths shorter than $2\pi/q^*$.

\begin{figure}
\includegraphics[width=0.45\textwidth,angle=0.0]{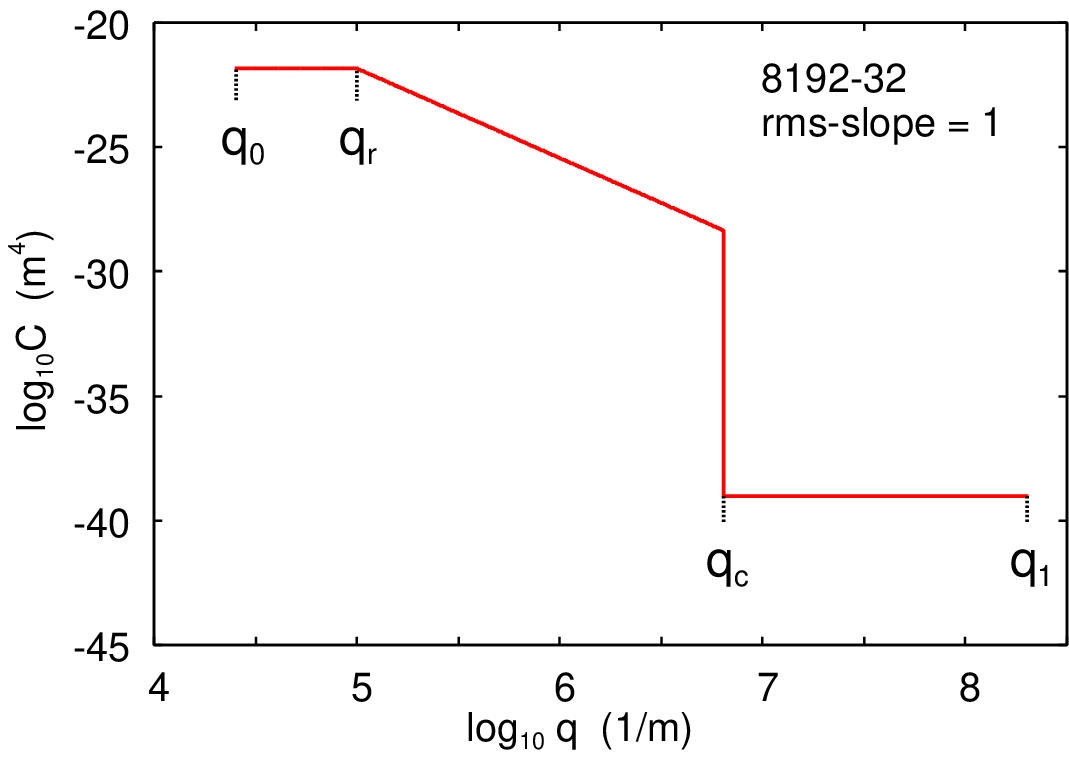}
\caption{\label{1logq.2logC.8196.32.eps}
Surface roughness power spectrum as a function of wavenumber (log–log scale) for the 8192--32 surface. 
The rms slope is $\xi = 1$, and the ratio between the largest and smallest wavenumber is $q_1/q_0 = 8192$. 
The width, $q_1/q_{\rm c}$, of the high-wavenumber cut-off region is $32$. 
The lowest and highest wavenumbers, $q_0$ and $q_1$, together with the roll-off and cut-off wavenumbers, $q_{\rm r}$ and $q_{\rm c}$, are indicated.
}
\end{figure}

To illustrate the procedure for deriving $C_{\rm pl}(q)$, we will use the surface denoted by 8196--32 in \cite{PRE}, where the ``width'' of the cut-off region is $q_1/q_{\rm c} = 32$ and the total ``length'' of the power spectrum is $q_1/q_0 = 8192$. Its power spectrum is depicted in Fig.~\ref{1logq.2logC.8196.32.eps}. The original surface topography height data and the plastically deformed topographies, used to simulate the unloading procedures with the boundary element method (BEM), have been made openly available \cite{almqvist_persson_2025_surface_topographies}. The width of the \emph{high-wavenumber} cut-off region $q_{\rm c} < q < q_1$ (here $q_1/q_{\rm c} = 32$) is irrelevant for the theory, but for the numerical study (using the boundary element method), a long enough high-wavenumber cut-off region is required in order to obtain converged results (see below).

\begin{figure}
\includegraphics[width=0.45\textwidth,angle=0.0]{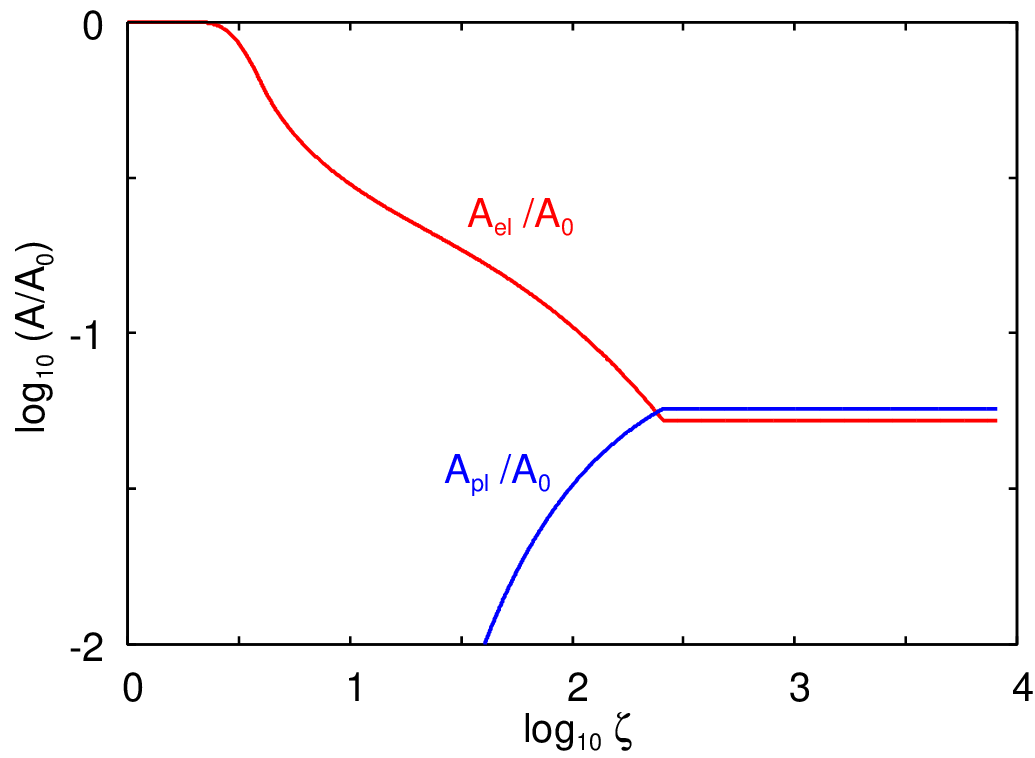}
\caption{\label{1logq.2Ael.Apl.8196.32.eps}
The elastic and plastic contact area as a function of the magnification (log-log scale), as predicted by the Persson's theory. 
The true contact areas are obtained at the highest magnification and are $A_{\rm pl}/A_0 = 0.0572$ and $A_{\rm el}/A_0 = 0.0526$. 
The total contact area $A/A_0 = (A_{\rm el} + A_{\rm pl})/A_0 =0.1099$. 
The results correspond Young's modulus $E = 250\ \mathrm{GPa}$, Poisson ratio $\nu = 0$, penetration hardness $\sigma_{\rm P} = 150\ \mathrm{GPa}$, and the applied squeezing pressure $\sigma_0 = 12.5\ \mathrm{GPa}$. 
The elastic energy reduction factor is $\gamma = 0.536$.}
\end{figure}

In all calculations we use a Young's modulus $E=250 \ {\rm GPa}$, Poisson ratio $\nu=0$, and penetration hardness $\sigma_{\rm P} = 150 \ {\rm GPa}$ (corresponding to a Yield stress in tension of $\approx 50 \ {\rm GPa}$). 
In the comparison between the Persson theory predictions and the numerical simulation results presented in Sec.~8, a range of applied pressures $\sigma_0$ is considered, but here we take $\sigma_0 = 12.50\ \mathrm{GPa}$. For these parameters and with the power spectrum shown in Fig.~\ref{1logq.2logC.8196.32.eps}, Fig.~\ref{1logq.2Ael.Apl.8196.32.eps} shows the elastic and plastic contact area as a function of the magnification (log-log scale), as predicted by (1) and (2).

\begin{figure}
\includegraphics[width=0.45\textwidth,angle=0.0]{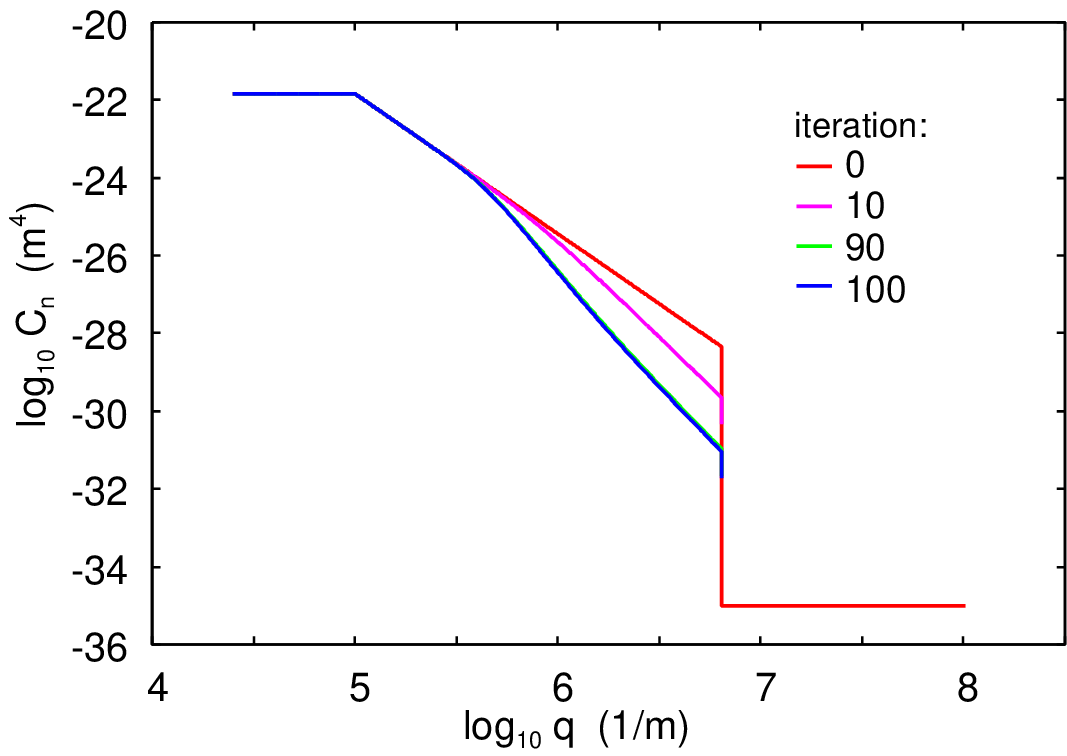}
\caption{\label{1logq.2logC.iterated.eps}
Power spectrum of the undeformed (original) surface (red), together with the plastically modified spectra after $N=10$ (pink), 90 (green), and 100 (blue) iterations. 
Results are for $\sigma_0/E^* = 0.05$.
}
\end{figure}

Figure~\ref{1logq.2logC.iterated.eps} shows the original power spectrum (red) and the plastically modified power spectrum after 10 (pink), 90 (green) and 100 (blue) iterations. The power spectra after 90 and 100 iterations are nearly identical, and in what follows we use for $C_{\rm pl}$ the $N=100$ result.

\begin{figure}
\includegraphics[width=0.45\textwidth,angle=0.0]{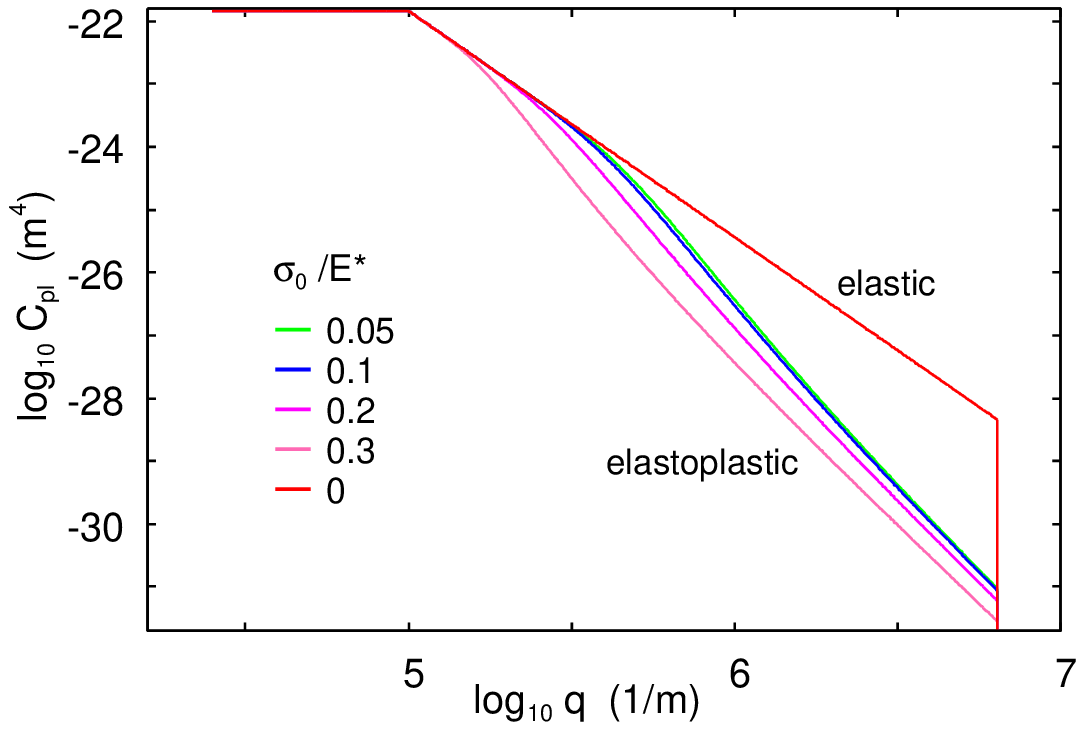}
\caption{\label{1logq.2logC.all.plastically.deformed.eps}
Power spectrum of the undeformed (original) surface (red), and the plastically modified spectrum after $N=100$ iterations for $\sigma_0/E^* = 0.05$ (green), 0.1 (blue), 0.2 (violet), and 0.3 (pink).
}
\end{figure}

Figure~\ref{1logq.2logC.all.plastically.deformed.eps} shows the original power spectrum (red) and the plastically modified power spectrum ($N=100$) for five values 0 (red), 0.05 (green), 0.1 (blue), 0.2 (violet) and 0.3 (pink), of the normalized applied squeezing pressure, defined as the ratio $\sigma_0/E^*$ between the applied squeezing pressure $\sigma_0$ and the effective elastic modulus $E^*$.


\vspace{0.3cm}
{\bf 6 Interfacial separation} 

Consider two solids with rough but nominally flat surfaces pressed together under a uniform external pressure $p$. As $p$ increases, the average surface separation, $\bar u$, decreases monotonically. 
We may therefore regard $\bar u (p)$ as a function of $p$, or equivalently $p(\bar u)$ as a function of $\bar u$. 
As the load increases, the asperities deform elastically and, for sufficiently large $p$, also plastically, and the elastic deformation energy (per unit surface area) $U_{\rm el}$ will be stored in the solid close to the interface. 
We can consider $U_{\rm el}$ as a function of either $\bar u$ or $p$, and since $U_{\rm el}$ equals the work performed by the external load in bringing the solids closer together, it follows that
$$
p(\bar u) = - \frac{d U_{\rm el}}{d \bar u} 
= - \frac{d U_{\rm el}}{dp} \, \frac{dp}{d \bar u}
$$
or
$$
d\bar u = -{1\over p} {d U_{\rm el} \over dp} dp. \eqno(6)
$$
As $p \rightarrow \infty$ then $\bar u \rightarrow 0$. 
Integrating (6), we get
$$
\bar u = -\int_p^\infty dp \ {1\over p} {d U_{\rm el} \over dp}.\eqno(7)
$$

The elastic energy can be written as
$$
U_{\rm el} = E^* {\pi \over 2} \int_{q_0}^{q_1} dq \ q^2 C(q) S(q,p),\eqno(8)
$$
where $q_1 = \zeta_1 q_0$ corresponds to the shortest wavelength  component in the roughness spectrum included at the highest magnification $\zeta_1$, and
$$
S(q,p) = \gamma + (1-\gamma )P(q,p),\eqno(9)
$$
where
$$
P(q,p) ={\rm erf} \left ({p\over 2G^{1/2}(q)} \right).\eqno(10)
$$
Using (7) and (8), one can calculate the dependence of $\bar u$ on the pressure $p$. Note that, replacing $q_1$ by $\zeta q_0$ yields $\bar u(\zeta)$ as a function of magnification, but here we focus on $\bar u$ at the highest magnification $\zeta_1$.

The contact stiffness is given by 
$$
K_0 = - {dp \over d\bar u}.\eqno(11)
$$
For small contact pressures $p$ (or large $\bar u$) the equations above give (see Refs.~\cite{PRL,Yang})
$$
p=p_0 e^{-\bar u/u_0},\eqno(12)
$$
where $p_0 = \beta E^*$ where $\beta$ and $u_0$ can be calculated from the surface roughness power spectrum. Typically, 
for self-affine fractal surfaces without a roll-off, see Ref.~\cite{PRL}, one finds $\beta \approx 0.4 q_0 h_{\rm rms}$ and $u_0 \approx 0.5 h_{\rm rms}$, where $h_{\rm rms}$ is the root-mean-square roughness amplitude. Combining (11) and (12) gives the contact stiffness for small pressures 
$$
K_0 = {p\over u_0}.\eqno(13)
$$
The theory above applies to infinitely large surfaces. An infinitely large surface with random roughness has a Gaussian distribution of asperity heights, and thus contains (infinitely many) infinitely high asperities. This implies that  $p$ is non-zero for arbitrarily large surface separation $\bar u$. 

For a finite system, however, the maximum asperity height is finite and depends on the particular surface realization. Consequently, the maximum height fluctuates from one realization to another, giving rise to what we denote as a \emph{finite-size effect}. These fluctuations diminish as the roll-off region of the power spectrum increases, since wider roll-offs enhance self-averaging. This property is useful in numerical studies to reduce statistical variation in the calculated contact properties. Also in the finite-size regime, the relation between $\bar u$ and $p$ can be obtained from (7), but with an elastic energy $U_{\rm el}$ determined as follows. For the details about the calculation, see 	\cite{pastewka2013_FiniteSize}. 

First, consider a rigid flat surface pressed against a nominally flat elastic body with a randomly rough surface. The longest wavelength roughness defines {\it macroasperities}. Although shorter-wavelength roughness sits on top of these, we first neglect it and treat the macroasperities as smooth. This implies that, for large enough average surface separation, the contact between the solids will occur at the highest macroasperity. 
The corresponding elastic energy (stored in the solid close to the interface), will be denoted $U_{0}$. It can be calculated using the Hertz contact theory and depends on the radius of curvature of the macroasperity $R_{0}$, which in turn is determined from the surface-roughness power spectrum $C(q)$.

Next, we take into account the shorter-wavelength roughness which on the macroasperity. This additional roughness contributes an elastic response that is, to leading order, analogous to the Hertzian contribution obtained from the macroasperity alone. The total elastic energy can therefore be written as $U_{\rm el} = U_{0} + U_{1}$, where $U_{1}$ represents the contribution from the smaller-scale roughness. Substituting this expression into (7) yields the relation $\bar u(p)$ in the finite-size regime. 
In~\cite{pastewka2013_FiniteSize}, it was assumed that the surface is a self-affine fractal, but the theory is valid for surfaces with an arbitrary surface roughness power spectrum, as is clear from the derivations in Appendix~B in \cite{pastewka2013_FiniteSize}.

If the surface roughness spectrum includes a long roll-off region, there exists a wide range of separations over which (12) holds. However, as the separation increases further, the solids eventually make contact only at the highest macroasperity. In \cite{pastewka2013_FiniteSize}, we derived an expression for the contact stiffness in the finite-size regime, which we denote as $K_1(p)$. Using the definition
$$
K_1 (p) = - {d p \over d\bar u},\eqno(14)
$$
gives us the increment in 
$$
d\bar u = - {1\over K_1(p)} dp.\eqno(15)
$$
If we assume the contact stiffness varies continuously across the crossover between the roll-off and finite-size regimes, then (13) is valid for $p>p^*$ and (14) for $p<p^*$, where
$$
K_1 (p^*) = {p^* \over u_0}.\eqno(16)
$$
Integrating (15) gives
$$
\bar u-u^* = \int_p^{p^*} dp \ {1\over K_1(p)},\eqno(17)
$$  
where, using (12), $u^* = u_0 {\rm ln} (p_0/p^*)$.

Although, the study of the finite-size effect in \cite{pastewka2013_FiniteSize} was focused on the contact stiffness, the relation between $\bar u$ and $p$ can be obtained either using (17) or (7). In the software we have developed, we first calculate the stiffness $K_1(p)$ and then obtain $\bar u(p)$ using (17). The implementation does not assume self-affine fractal surfaces; instead, the power spectrum $C(q)$ is provided numerically and may have an arbitrary form.

\vspace{0.3cm}
{\bf 7 Boundary element method} 

In this study, we employ the same boundary element method (BEM) as in \cite{PRE}. We simulate frictionless contact between an elastic--perfectly plastic half-space with a rough surface and a rigid flat body. 
The surface roughness is represented as a periodic height field, and deformations are computed in the Fourier space, via convolution with the Green's function of an elastic half-space. Plastic deformation is introduced through a local yield criterion: whenever the local contact pressure exceeds the penetration hardness $\sigma_{\rm P}$, the corresponding node is treated as plastically loaded and constrained to remain in contact without carrying additional load. 

The solver employs a relaxation-based iterative scheme with residual convergence to enforce global force balance. This BEM framework has been applied in several earlier works; see Refs.~\cite{Almqvist2007, Sahlin2010, Almqvist2011, metal1, PerezRafols2021, Kalliorinne2021, Kalliorinne2022, kalliorinne2023}.

\begin{figure}
\includegraphics[width=0.45\textwidth,angle=0.0,trim=8cm 0cm 8cm 0cm]{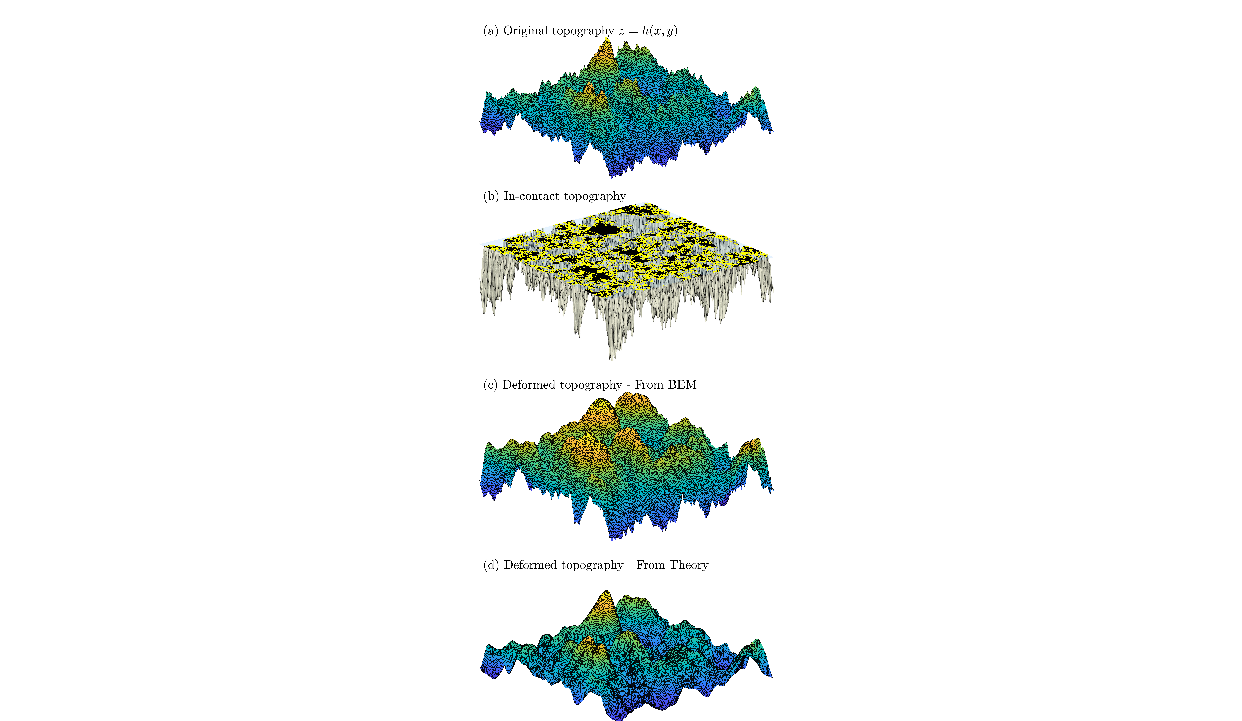}
\caption{\label{OriginalAnContact.eps}
(a) The original height topography, $z=h(x,y)$, of the 8192--32 surface, (b) the calculated in-contact topography for the elastoplastic solid pressed against a rigid flat surface (transparent), and (c) and (d) the plastically deformed topography after unloading from the BEM and theory, respectively. The results in (b) and (c) were obtained with the boundary element method (BEM), modeling the solid as an elastic--perfectly plastic half-space with Young's modulus $E = 250\ \mathrm{GPa}$, Poisson ratio $\nu = 0$, penetration hardness $\sigma_{\rm P} = 150\ \mathrm{GPa}$, and the applied squeezing pressure $\sigma_0 = 50\ \mathrm{GPa}$. The calculated relative elastic contact area is $A_{\rm el}/A_0 = 0.1540$ (yellow), and the plastic contact area $A_{\rm pl}/A_0 = 0.2567$ (black), and total contact area $A/A_0 = 0.4107$.
}
\end{figure}



\begin{figure}
\includegraphics[width=0.47\textwidth,angle=0.0]{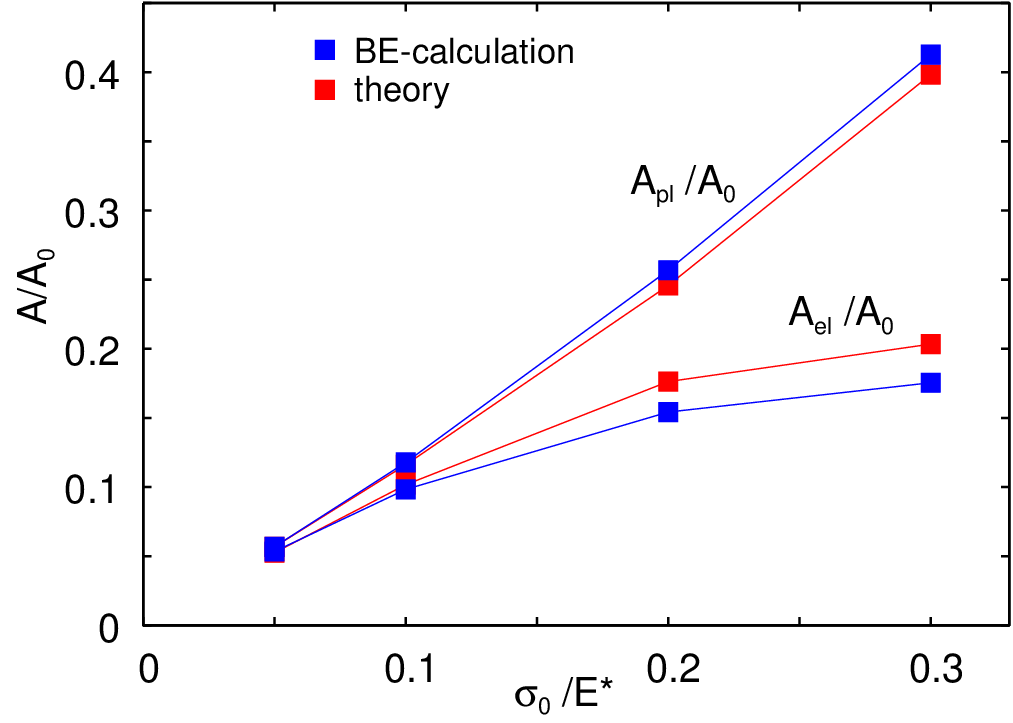}
\caption{\label{1pOverE.2AplAel.eps}
The relative elastic ($A_{\rm el}/A_0$) and plastic ($A_{\rm pl}/A_0$) contact areas as functions of the normalized applied squeezing pressure, defined as the ratio $\sigma_0/E^*$ between the applied squeezing pressure $\sigma_0$ and the effective elastic modulus $E^*$. 
The results were obtained using the same 8192--32 surface, and material parameters as for the results presented in Fig.~\ref{OriginalAnContact.eps}.
}

\end{figure}

\begin{figure}
\includegraphics[width=0.47\textwidth,angle=0.0]{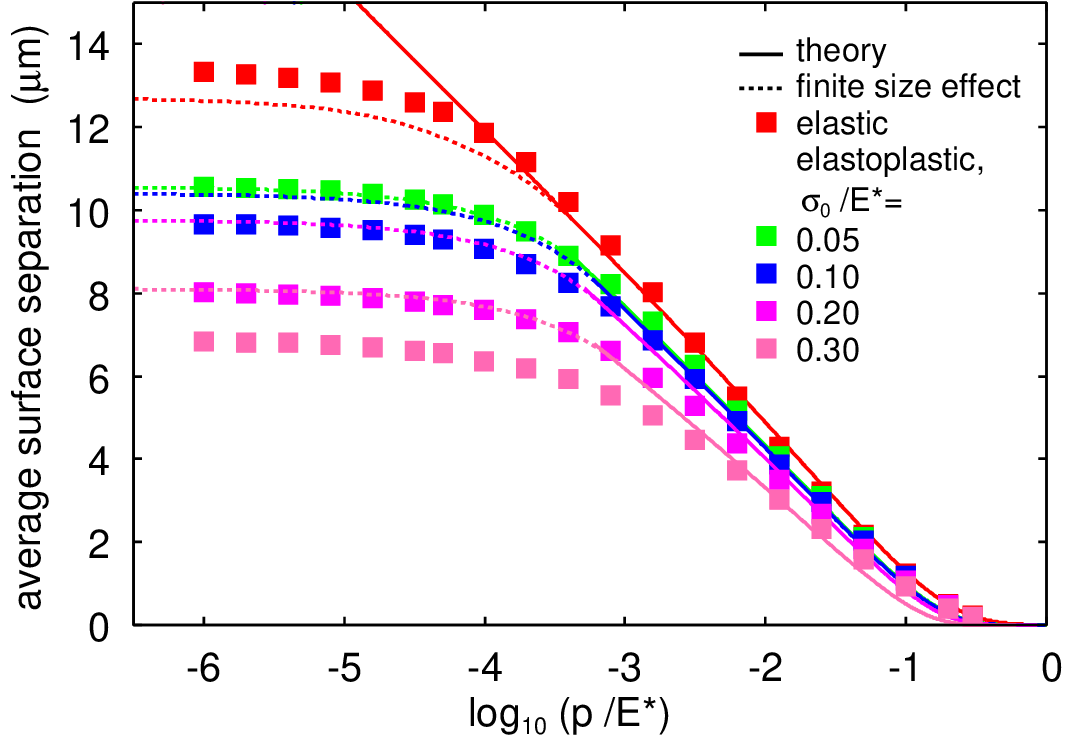}
\caption{\label{1logPoverE.2separation.eps}
The average surface separation $\bar u$ as a function of the logarithm of the normalized squeezing pressure $p/E^*$. For the 8192--32 system. The red squares are the results of the BEM numerical calculations without plastic deformation. The green, blue, pink and violet squares are the numerical BEM results, starting from the surfaces plastically deformed by $\sigma_0/E^*=0.05$, $0.10$, $0.20$, and $0.30$, respectively. The solid lines represent the Persson theory prediction neglecting the finite-size effect, and the dashed lines represent the theory prediction including the finite-size effect. 
}
\end{figure}

\vspace{0.3cm}
{\bf 8 Comparison of theory with numerical results} 

In this study, we use the 8192--32 surface introduced in \cite{PRE}, where the ``width'' of the cut-off region is $q_1/q_{\rm c} = 32$ and the total ``length'' of the power spectrum is $q_1/q_0 = 8192$. This randomly rough surface was generated by superposing plane waves with random phases. We also use the same material parameters as in \cite{PRE}, but in the present analysis we consider several values of the applied squeezing pressure $\sigma_0$. In \cite{PRE} we used three surfaces, 2048--8, 4096--16, and 8192--32, but here we focus on the largest system since theory effectively corresponds to $q_1/q_{\rm c} = \infty$.

The 8192--32 surface exhibits self-affine fractal roughness in the wavenumber interval $1 \times 10^5 \ {\rm m}^{-1} < q < 64 \times 10^5 \ {\rm m}^{-1}$, with a fractal dimension $D_{\rm f} = 2.2$ (corresponding to a Hurst exponent $H = 0.8$). Its power spectrum contains a short roll-off region in the range $2.5 \times 10^4 \ {\rm m}^{-1} < q < 1 \times 10^5 \ {\rm m}^{-1}$, which enhances self-averaging. Moreover, since most engineering surfaces are designed to be smooth at the macroscopic scale, they also have a roll-off region in the power spectra, see Ref. \cite{RollOff}.

When comparing numerical simulations with continuum mechanics theories, it is crucial to have enough grid points within the shortest roughness wavelength. This requirement can be met by using surfaces which are generated from power spectra that have a wide enough cut-off region for the large wavenumbers, as illustrated in Fig.~\ref{1logq.2logC.8196.32.eps} for the 8192--32 surface. The surface height topography, $z=h(x,y)$, has the rms-slope $\xi =1$.

For convenience, we repeat here the material parameters given in Sec.~5, which are the same as those in \cite{PRE}: Young's modulus $E = 250\ \mathrm{GPa}$, Poisson ratio $\nu = 0$, and penetration hardness $\sigma_{\rm P} = 150\ \mathrm{GPa}$. 

In elastic contact mechanics, as long as the real contact area $A<<A_0$, it scales linearly with the normalized applied squeezing pressure, defined as the ratio $\sigma_0/E^*$ between the applied squeezing pressure $\sigma_0$ and the effective elastic modulus $E^*$, according to
$$ 
\frac{A}{A_0} \approx \frac{2}{\xi}\frac{\sigma_0}{E^*}, \eqno(18)
$$
where $E^*=E/(1-\nu^2)$ is the effective modulus and $\xi$ the rms slope. In the present case, $E^*=E= 250 \ {\rm GPa}$ and $\xi=1$. Hence, for an elastic contact and the applied squeezing pressure 
$\sigma_0 = 12.5 \ {\rm GPa}$ (that was used in \cite{PRE}), one finds $A/A_0 = 0.1$. In the present study, in addition to $\sigma_0 = 12.5 \ {\rm GPa}$, we also consider $\sigma_0 = 25$, $50$, and $75 \ {\rm GPa}$, corresponding to $\sigma_0/E^* = 0.05$, $0.1$, $0.2$, and $0.3$. The resulting plastically deformed surfaces employed in the unloading simulations are openly available; see Ref.~\cite{almqvist_persson_2025_surface_topographies}.

Figure~\ref{OriginalAnContact.eps} shows (a) the original height topography $z=h(x,y)$ of the 8192--32 surface, (b) the in-contact topography obtained using the BEM for the surface pressed against a rigid flat body, with the solid modelled as an elastic–perfectly plastic half-space and an applied squeezing pressure of $\sigma_0 = 50\ \mathrm{GPa}$, and (c) the plastically deformed topography after unloading from $\sigma_0 = 50\ \mathrm{GPa}$.
Figure~\ref{OriginalAnContact.eps}(d) shows $h(x,y)$ for the plastically smoothed power spectrum generated using the same sequence of random numbers as for the original surface.
Note that the highest asperity in the theory (d) is sharper than in the  plastically deformed surface obtained with the BEM in (c).
The total relative contact area is $A/A_0 = 0.4107$, is composed of the elastic contribution $A_{\rm el}/A_0 = 0.1540$ (yellow) and the plastic contribution $A_{\rm pl}/A_0 = 0.2567$ (black).

Figure~\ref{1pOverE.2AplAel.eps} shows the relative elastic ($A_{\rm el}/A_0$) and plastic ($A_{\rm pl}/A_0$) contact areas as functions of the normalized applied squeezing pressure $\sigma_0/E^*$. The theoretical predictions are in close quantitative agreement with the numerical BEM results.

Figure~\ref{1logPoverE.2separation.eps} shows the average surface separation $\bar u$ as a function of the logarithm of the normalized squeezing pressure $\sigma_0/E^*$. The red squares denote the BEM results without plastic deformation, while the green, blue, pink and violet squares are the numerical BEM results, starting from the surfaces plastically deformed by $\sigma_0/E^*=0.05$, $0.10$, $0.20$, and $0.30$, respectively. The solid lines represent the Persson theory predictions neglecting finite-size effects, and the dashed lines represent the predictions including finite-size effects. Overall, the theoretical predictions are consistent with the BEM results and highlight the influence of finite-size effects.



\begin{figure}
	\includegraphics[width=0.47\textwidth,angle=0.0]{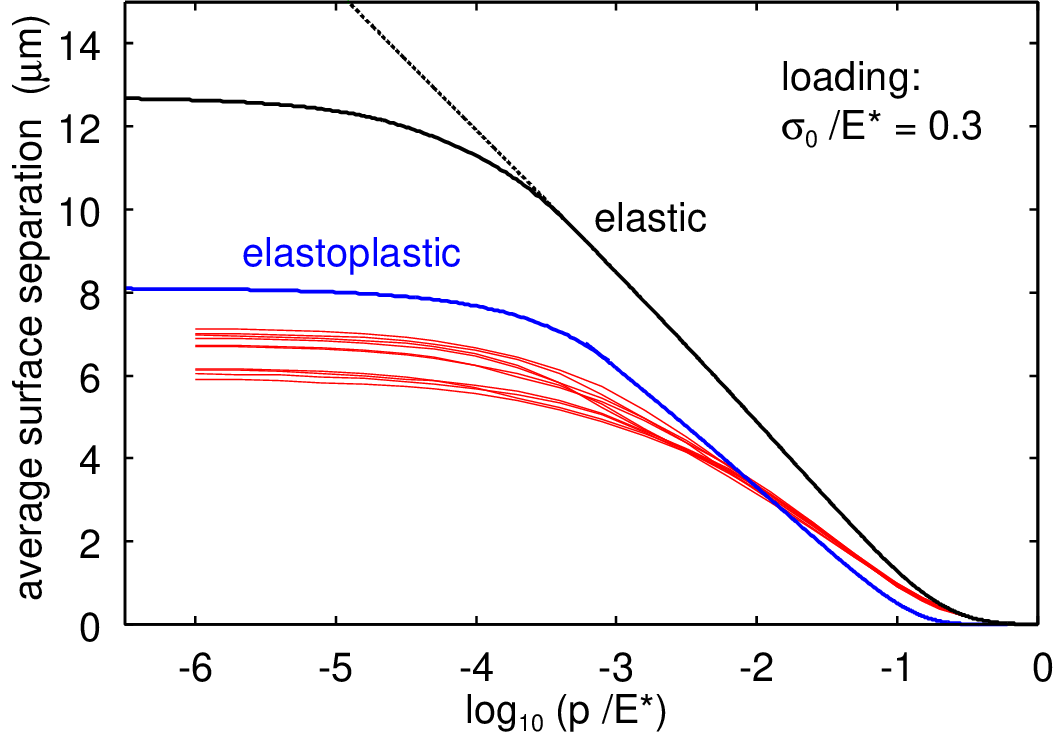}
	\caption{\label{1logpOverE.2baru.all.realizations.eps}
		The surface separation $\bar u$ as a function of the logarithm of the normalized squeezing pressure $p/E^*$. 
		The surfaces was first loaded against the flat surface with the normalized applied squeezing pressure $\sigma_0/E^*=0.3$ and thereafter the pressure was lowered.
		The thin lines are the BEM numerical calculations for 10 different realizations of the 8192--32 system.  
		The blue solid line represent the Persson theory prediction, and 
		the black line the theory prediction assuming no plastic deformation.
	}
\end{figure}

The numerical BEM results correspond to a single realization of the surface, whereas the theory represents ensemble-averaged behavior. Since the height and shape of the highest asperity vary from one realization to another, and since at low squeezing pressures the rigid flat surface makes contact only with the highest macroasperity, relatively large finite-size fluctuations in the $\bar u(p)$ relation are expected. To test this we generated 10 different realizations of the 8192--32 surface using different seeds for the random number generator. The different surfaces was loaded against the flat surface with $\sigma_0/E^* = 0.3 \ {\rm MPa}$ after which the pressure was reduced.

The thin lines in Fig.~\ref{1logpOverE.2baru.all.realizations.eps} shows the relation between the average surface separation and the applied pressure as obtained from the numerical simulations during retraction. The black line is the theory result assuming elastic contact (infinite penetration hardness) and the blue line the theory result using the same penetration hardness as in the simulations. (The blue line is the same result as as the pink line in Fig.~\ref{1logPoverE.2separation.eps}.) 

Note that the theory gives larger average separation than the simulation model. This may be related to the different smoothing procedures 
and it is not clear which smoothing procedure gives the most accurate description. In the simulation model asperities can be fully flattened but this is not the case in reality: when asperities deform plastically the stress field approach that of a hydrostatic field and asperities cannot be fully flattened (see Ref. \cite{Tiwari2020_PlasticDefRoughMetallicSurface_TribLett} for experimental results illustrating this). This correspond to a larger interfacial separation, as observed in the theory, for a given applied pressure.

By contrast, the real contact area is governed primarily by the short-wavelength roughness and is therefore much less sensitive to the particular surface realization.
The simulation results for the 10 different realizations gives curves which cannot be distinguished in a graph and is therefore not shown here.

\vspace{0.3cm}
{\bf 9 Discussion}

We have compared Persson's multiscale contact mechanics predictions for the relation $\bar u(p)$ with numerical BEM simulations of an elastic–perfectly plastic half-space. In the simulations, the solid deforms elastically until the local stress reaches the penetration hardness $\sigma_{\rm P}$, after which plastic flow occurs without work (strain) hardening. The comparison shows that the procedure---where plastic smoothing is represented by an effective roughness power spectrum within the Persson framework---captures the overall trends observed in the simulations.

The boundary element method use deterministic surfaces whereas the theory describes an ensemble average. At low squeezing pressures, in the finite-size region, the average surface separation $\bar u$ depends mainly on the long wavelength roughness, which differ for different realizations of the surface. As a result $\bar u$ for low contact pressures exhibit large finite-size fluctuations (see Fig.~\ref{1logpOverE.2baru.all.realizations.eps}).
In contrast, the real contact area depends mainly on short-wavelength roughness and is therefore much  less sensitive to finite-size variability. The elastic and plastic contact area predicted by the theory is in the close quantitative agreement with the BEM result (see Fig.~\ref{1pOverE.2AplAel.eps}).

The finite-size treatment within Persson's theory assumes that macroasperities can be represented as spherical bumps whose elastic stiffness follows Hertz theory. When plastic deformation extends to the macroasperity scale, this assumption becomes less accurate, since plastic flattening increases the elastic stiffness upon reloading. Numerical indentation studies confirm that even a small flat region at the top of a spherical bump increases the elastic stiffness $K = dF/d\delta$, see Ref.~\cite{korsunsky2001FlatPunch}. In our simulations at $\sigma_0/E^* = 0.3$, increasing $K$ by a factor of two in the finite-size model restores agreement with the BEM results (Fig.~\ref{1logPoverE.2separation.eps}).

The numerical results presented in our study could be tested experimentally by squeezing rectangular metal blocks against flat substrate surfaces and measure the movement of the top surface as a function of the normal force. Such studied must take into account the elastic deformations of the solids and would probably only be possible for systems with ``large'' interfacial roughness. Similar studies has been  used for rubber squeezed against different surfaces, from which the average surface separation was deduced \cite{Rubber1, Rubber2}.  Another test is fluid leakage measurements, but the leakage rate depends not directly on the average surface separation, but on the surface separation at the most narrow constriction along the biggest open fluid flow channel \cite{Leakage}. Other indirect measurements would be measurements of the contact stiffness \cite{Stiffness}.

These findings emphasize both the strengths and limitations of the present procedure. 
The effective-spectrum approach provides a simple and robust way to incorporate plastic smoothing into Persson's theory, 
yielding useful predictions for the interfacial separation across a broad load range. 
The theory also predict the contact area, which does not depend on the smoothed surface roughness profile, in good agreement with the BEM predictions.
At the same time, both the BEM and the analytic theory should be extended to include a more realistic description of the plastic flow process as described in Sec.~2.
From an applications perspective, the results highlight that plastic adaptation leads to reduced interfacial separation---an effect of practical relevance for sealing, leakage, and wear.

\vspace{0.3cm}
{\bf 10 Summary and conclusion} 

We have proposed a procedure within Persson's multiscale contact mechanics theory to describe elastoplastic contacts. The method accounts for plastic smoothing by applying the elastic formulation to an effective power spectrum of the plastically deformed surface. Comparisons with BEM simulations show close agreement for elastic and plastic contact areas and good agreement for the average surface separation, with deviations traceable to finite-size fluctuations and plastic macroasperity deformation. The approach captures the reduction of separation upon reloading---relevant to sealing and leakage---and offers a basis for incorporating plastic stiffening at large loads.

%
%
%


%

\end{document}